\documentclass[aps,final,notitlepage,oneside,twocolumn,nobibnotes,nofootinbib,
superscriptaddress,noshowpacs,centertags]{revtex4-1}

\usepackage[english]{babel}
\usepackage{graphicx}
\usepackage{latexsym}
\usepackage{amssymb}
\usepackage{amsmath}
\usepackage{float}

\begin{document}

\title{Physical laboratory at the center of the Galaxy}
\author{V.\,I.\:Dokuchaev}\thanks{e-mail: dokuchaev@lngs.infn.it}
\affiliation{Institute for Nuclear Research of the Russian Academy of
Sciences \\ 60th October Anniversary Prospect 7a, 117312 Moscow, Russia }
\affiliation{National Research Nuclear University MEPhI
(Moscow Engineering Physics Institute),
Kashirskoe shosse 31, 115409 Moscow, Russian Federation}
\author{Yu.\,N.\:Eroshenko}\thanks{e-mail: eroshenko@inr.ac.ru}
\affiliation{Institute for Nuclear Research of the Russian Academy of
Sciences \\ 60th October Anniversary Prospect 7a, 117312 Moscow, Russia }

\date{\today}

\begin{abstract}
We review the physical processes that occur at the center of the Galaxy and that are related to the supermassive black hole Sgr A* residing there. The discovery of high-velocity S0 stars orbiting Sgr A* for the first time allowed measuring the mass of this supermassive black hole, the closest one to us, with a 10\% accuracy, with the result $M_h=(4.1\pm0.4)\times 10^6M_\odot$. Further monitoring can potentially discover the Newtonian precession of the S0 star orbits in the gravitational field of the black hole due to invisible distributed matter. This will yield the ``weight'' of the elusive dark matter concentrated there and provide new information for the identification of dark matter particles. The weak accretion activity of the ``dormant quasar'' at the Galactic center occasionally shows up as quasiperiodic X-ray and near-IR oscillations with mean periods of $11$ and $19$ min. These oscillations can possibly be interpreted as related to the rotation frequency of the Sgr A* event horizon and to the latitude oscillations of hot plasma spots in the accretion disk. Both these frequencies depend only on the black hole gravitational field and not on the accretion model. Using this interpretation yields quite the accurate values for both the mass $M_h$ and the spin $a$ (Kerr rotation parameter) of Sgr A*: $M_h=(4.2\pm0.2)\times 10^6M_\odot$ and $a=0.65\pm0.05$.
\end{abstract}

\maketitle 

\tableofcontents



\section{Introduction}

The Galactic center is of utmost interest owing to the presence of the central supermassive black hole Sagittarius A* ( Sgr~A*), the closest supermassive BH (SMBH), at a distance of 8.5~kpc with a mass of $4\times10^6M_\odot$. Clearly, the identification of the radio source Sgr~A* (and similar objects in other galaxies) with a black hole is solely based on the fact that it is difficult to consistently associate the bulk of observational data with other astronomical objects \cite{1,2,3,4,5,6,7,8,9,10,11,12}. To make indirect experiments scientifically rigorous and compelling, sufficient completeness, consistency, and uniqueness of a large amount of observational data are required \cite{13}.

In the case of black holes, the definitive experiment is the discovery of the event horizon, which simultaneously is an important test of Einstein's general relativity (GR) in the strong-field regime. The discovery of the event horizon of the central Galactic BH will possibly happen already in this decade.

The SMBH Sgr~A* qualifies as a ``dormant'' quasar due to the low current activity across practically all the electromagnetic spectrum except the radio band. The low luminosity of Sgr~A* additionally complicates its study. In the optical band, the Galactic center region is unavailable for observations due to the high optical depth caused by light scattering and absorption by the interstellar dust (mostly consisting of
carbon) in the Galactic disk. Nevertheless, the dust turns out to be sufficiently transparent in the near infrared (IR) band inside the so-called Baade window, at the edge of which the Galactic center is located. Due to this lucky fact, observations of Sgr~A* and its surroundings by ground-based telescopes are possible.

The Galactic center is also observed in the radio, X-ray, and gamma-ray bands. In the near-IR band, as well as in X-rays, Sgr~A* shows up as a transient source with rare outbursts; in gamma-rays, the Galactic center has not yet been reliably identified. The mean bolometric luminosity of Sgr~A* is $\lesssim10^{36}$~erg~s$^{-1}$. Various models have been elaborated to interpret the observed emission from Sgr~A* (see, e.\,g., \cite{14,15,16,17,18,19,20}).

The measurement of the mass of the supermassive black hole in the Galactic center with a record high accuracy up to 10\% is an important scientific achievement of the 21st century. It was made possible thanks to the analysis of multi-year IR observations of the high-velocity S0 stars that move around Sgr~A* in eccentric orbits \cite{21,22,23,24,25,26,27,28,29}. The S0-2 star moving around the BH in an orbit with the eccentricity $e=0.89$, major semi-axis $a=5$~mpc, orbital period $P=16$~yrs, and periastron velocity $v=1500$~km~s$^{-1}$ is studied best. The highest velocity $v=1.2\times10^4$~km~s$^{-1}$, is
observed for the S0-16 star. The star S0-102 has the shortest orbital period, $P=11.5$~yrs.

Measurement of the Keplerian orbital parameters of the S0 stars over more than 20 years allowed the first
precise determination of the mass of Sgr~A*: $M_h=(4.1\pm0.4)\times10^6M_\odot$ \cite{23,24,25,26,26}. The stars S0-2 and S0-16 have been observed over more than one orbital period, which will further improve the black hole mass measurement.

Relativistic post-Newtonian effects in the motion of the known S0 stars are small and cannot be measured yet \cite{30,31,32}. However, the discovery of even faster S0 stars in the Galactic center is quite possible. Of great interest would be the detection of a pulsar moving around the BH Sgr~A* in a short-period orbit.

The mass and angular momentum of the Galactic central BH can be inferred from observations of quasiperiodic
oscillations. This method is applicable due to the relatively low luminosity of Sgr~A*, at which the plasma surrounding the BH is transparent down to the event horizon. Hot spots in the inner accretion disk or hot plasma blobs in nonequatorial orbits around the BH could produce such quasiperiodic oscillations. The signal from these hot blobs should be modulated with two characteristic frequencies: the rotational frequency of the BH event horizon and the frequency of latitudinal oscillations of the blob orbit. Neither of these frequencies depends on the accretion model, and both are fully determined by the BH gravitational field.

The interpretation of the quasiperiodic oscillations observed from the Galactic supermassive BH in the X-ray
and IR bands by the two characteristic frequency methods yields the most exact estimate of the mass $M_h$ and the spin $a$ (Kerr parameter) of Sgr~A*: $M_h=(4.2\pm0.2)\times 10^6M_\odot$ and $a=0.65\pm0.05$ \cite{33}.

Another fundamental problem discussed with regard to the central BH is the nature of dark matter, which is a major astrophysical enigma. If dark matter consists of elementary particles that can annihilate or decay, gamma-ray emission can be generated. Currently, an extensive search for gamma-ray emission from dark matter is being carried out by the Fermi-LAT (Fermi Large Area Telescope) gamma-ray telescope.

The Galactic center is one of the most suitable sites to search for dark mater signals, despite the presence of strong backgrounds from the Galactic disk. Analytic models and numerical simulations of the dark mater halo suggest that the halo density strongly increases toward the center. Therefore, the annihilation signal from the Galactic center is expected to be quite significant \cite{34}. In addition, the formation of an
additional dark matter peak around the central supermassive BH is predicted, which should appear as a bright point-like source in gamma rays. The Fermi-LAT space telescope reported an intriguing gamma-ray excess from the Galactic center with an angular size of several degrees, which cannot be explained by usual astrophysical sources but could be due to dark matter annihilation \cite{35,36,37,38,39,40,41,42,43,44}. This result is still unreliable and requires additional checks. Presently, the search for the annihilation dark matter gamma rays is one of the ``hot'' points in astrophysics. The problem could be solved by the planned Russian Gamma-400 gamma-ray space observatory \cite{45}.


\section{{\em Experimentum crucis} for discovery of the event horizon}

Without a doubt, great galactographic discoveries will be made in the 21st century primarily due to GR tests in the natural physical laboratory at the Galactic center. Possibly, already in the next decade, the event horizon of the Sgr~A* BH will be discovered. With the mass $M_h=4.2\times10^6M_\odot$, the linear size of the BH event horizon in the case of extremely rapid spin is $r_g= GM_h/c^2 \simeq 6.2 \times10^{11}$~cm, which is about nine solar radii. From the distance $d_c\simeq8$~kpc to the Galactic center, an object of this size would have the angular diameter $\phi_g=2 r_g/d_c \sim{10^{-5}}''$. To resolve it, an instrument with an
angular resolution of a few micro arc seconds is needed. Currently, only very long baseline radio interferometry (VLBI) in the millimeter and submillimeter bands enables the necessary micro-arcsecond resolution. 

Since 2007, a substantial international collaboration has worked out the project called the Event Horizon Telescope (EHT), which is primarily aimed at identifying the event horizon of the supermassive BH in the Galactic center. The next prospective candidate is the SMBH at the center of the nearby giant elliptical galaxy M87 \cite{48}. In that galaxy, however, the plasma surrounding the BH could be nontransparent due to a
high accretion rate onto the BH. The EHT project includes an array of radio telescopes located on different continents, which will operate as a single VLBI radio interferometer at the frequency of 230~GHz
(wavelength 1.3~mm). By 2020, the construction of a VLBI array including more than 13 antennas is planned, which will be capable of resolving the BH shadow illuminated either by background sources \cite{49,50,51,52,53,54,55,56,57} or by the surrounding accretion disk \cite{58,59,60,61,62,63}.

\begin{figure}
	\begin{center}
\includegraphics[angle=0,width=0.45\textwidth]{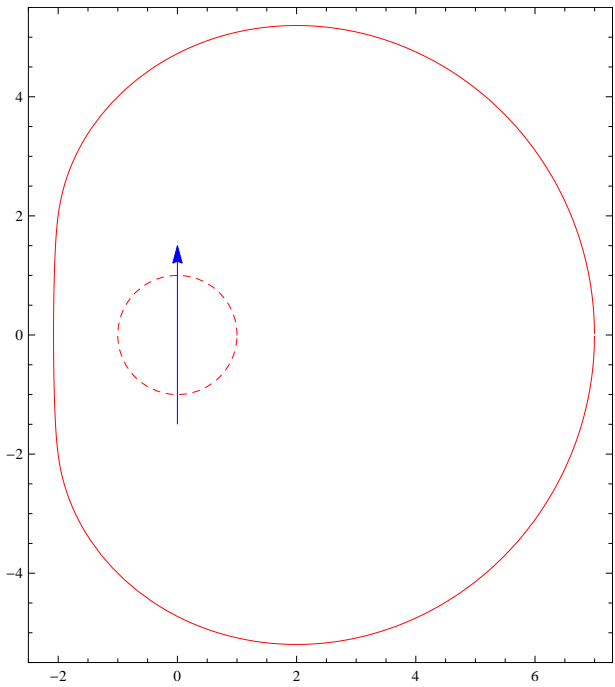}
	\end{center}
\caption{Shape of the external boundary of the shadow from an extreme Kerr BH against a bright background (in the linear units $GM_h/c^2$). The dashed circle corresponds to the BH horizon with the radius $r_g=GM_h/c^2$. The arrow shows the BH spin direction.}
	\label{BHshadow}
\end{figure}

In Russia, the space observatory Millimetron (the Spectrum-M project) is under development, which will include a radio telescope in the millimeter and submillimeter bands placed near the Lagrangian point L2 of the Earth-Sun system, where a nanoarcsecond angular resolution may be achieved in the radio interferometer regime \cite{64}. With such a high angular resolution, the BH ``shadow'' could be resolved. The shape of the outer boundary of the shadow from the extreme Kerr BH against a bright background is shown in Fig.~\ref{BHshadow}.

The EHT project will undoubtedly be an extremely important experiment of the 21st century, which will allow testing the existence of BHs and hence GR in the strong-field regime. In the future, a detailed measurement of the BH shadow form and hence the BH horizon will enable probing the effects of dark energy on BHs \cite{65,66}, testing different GR modifications \cite{67,68,69}, and, possibly, identifying wormholes \cite{70,71}.

\section{Observational appearances of the central Galactic black hole}
\label{observations}

As noted in the Introduction, direct IR observations of the so-called S0 stars moving near the Galactic center in eccentric orbits provide the strongest evidence of the presence of an SMBH at the Galactic center. In \cite{72}, the characteristic radial dependence of the stellar velocity dispersion, $\langle  v^2\rangle\propto1/r$, was derived for the first time from spectroscopic observations and measurements of proper motions of the stars, which corresponds to a Keplerian motion of the stars around a very
massive object. By 1996, the mass of the object was estimated to be $(2.45\pm0.4)\times10^6M_\odot$. According to modern measurements of these stars, the mass $M_h\sim4\times10^6M_\odot$ is enclosed within a sphere with the radius $\sim6\times10^{-4}$~pc. Such a very compact massive object can be identified only with a black
hole.

Recent VLBI data in the centimeter band suggest the presence of an emitting region around Sgr~A* approximately
$\sim10^{-3}{}''$ in size, which is apparently related to an accretion disk \cite{73}. A joint analysis of VLA (Very Large Array) interferometry at centimeter wavelengths and ALMA (Atacama Large Millimeter Array) observations at 100~GHz points to the presence of collimated outflows from Sgr~A* with moderate relativistic velocities $\sim0.5c$ \cite{74}.

The IR and X-ray activity of Sgr~A* is very low; it appears as a series of very rare and short-duration outbursts, which are interpreted in terms of accretion of gas clouds, planets \cite{75}, comets, or asteroids falling onto the BH. Nevertheless, during these rare outbursts, quasiperiodic oscillations in the IR \cite{76} and X-ray \cite{77} bands were discovered. The physical origin of these oscillations is discussed in Section~\ref{QPO}.

The SMBH in the Galactic center is `dormant' because of a low rate of tidal disruption of stars and a small accretion rate of the ambient gas. The rate of tidal disruption of stars by a $\dot N_{\rm fin}$, under the condition $r_{\rm cr}<r_h$ (see formula (\ref{cr}) in Section 4) can be represented in the form \cite{78,79,80,81}
\begin{equation}
\dot M_h\simeq\frac{2,7\times10^3}{\Lambda_{\rm cr}}\left(\frac{M_h}{Nm_*}\right)^3
\left(\frac{r_{\rm cr}}{r_h}\right)\frac{Nm_*}{t_{\rm E}},
 \label{Nfin}
\end{equation}
where $t_{\rm E}$ is the local relaxation time, $\Lambda_{\rm cr}=\ln(r_{\rm cr}/r_t)$. At the Galactic center, the tidal disruption rate $\dot M_h \sim10^{-4}M_\odot$~yr$^{-1}$ coincides within an order
of magnitude with the dark matter accretion rate from the Galactic halo \cite{82,83,84,85}. With such an accretion rate, the Sgr~A* BH could have grown to its present mass over the galaxy lifetime. The SMBH evolution in galactic centers due to tidal disruption of stars and dark matter accretion has been studied
for many years (also see, e.\,g., \cite{86,87,88}).

Although BHs in galactic centers are small in size, their powerful gravitational field can affect giant structures and processes on galactic scales and beyond. For example, the accretion luminosity of SMBHs in quasars could reionize all the gas in the Universe \cite{89}. In active galaxies, central BHs are
powerful emitters in various bands, and also produce relativistic jets starting near the BH gravitational radius and propagating far beyond the galaxies \cite{90}.

During each tidal disruption event, occurring once every $10^4$ years, a short-lived accretion disk is formed around the BH for about 5-10~years. During this time interval, the ``dormant'' quasar at the Galactic center ``wakes up'', its Eddington luminosity becomes comparable to that of all stars in the galaxy, and relativistic jets are generated perpendicular to the Galactic disk \cite{91,92,93,94}. The past activity of Sgr~A* could be traced back by recent observations of the diffuse iron line in 6.4~keV emission \cite{95} and hydrogen
ionization in the central 1~pc region \cite{96,97}.

Two giant diffuse gamma-ray ``bubbles'' discovered by the Fermi satellite symmetrically on both sides of the Galactic disk plane \cite{98,99} may be a long-term imprint of this activity. Fermi bubbles might be formed due to matter outflow from tidally disrupted stars flying close to the central BH \cite{100}. The
gamma-ray luminosity of the Fermi bubbles is $4\times10^{37}$~erg~s$^{-1}$ \cite{101}. The bubbles have a size of up to 10 kpc and have a structure similar to the nonthermal microwave haze discovered by the WMAP (Wilkinson Microwave Anisotropy Probe) satellite \cite{102,103}, as well as to the extended X-ray
emission lobes discovered by the ROSAT (from German ROSAT satellite) telescope \cite{104}. The features of gamma bubbles are consistent with the model of multiple energy injections from the Galactic center due to the central BH activity \cite{105,106}. However, there is an alternative model explaining the energy pumping into the gamma bubbles by the energetic winds of protons and heavy ions from star formation inside the inner region about 200~pc in size \cite{107,108}.


\section{Distribution of stars in the Galactic center}

The central star cluster \cite{29,109,110} comprises the mass $M_*\sim10^7M_\odot$ within $R_*\sim1$~pc. The corresponding virial velocity of stars in this cluster is $v_{\rm vir}\sim10^2$~km~s$^{-1}$, and their number density is very high, $n_*\sim10^7$~pc$^{-3}$, which is eight orders of magnitude higher than in the solar
vicinity, $0.12$~pc$^{-3}$. The radial density distribution found in \cite{109} from adaptive optical observations in the near IR using the VLT (Very Large Telescope) at ESO (European Southern Observatory) has the form $\rho_*(r)= 1.2\times10^6(r/0.4\mbox{~пк})^{-\alpha}M_\odot$~pc$^{-3}$, where $\alpha=1.4\pm0.1$ for $0.004$~pc~$<r<0.4$~pc and $\alpha=2.0\pm0.1$ for $r>0.4$~pc.

In the central cluster, old stars with enhanced metal abundance compared to the halo stars dominate. However,
among the brightest stars, there is a sizeable fraction of young stars with different metallicities. It is currently unclear whether the young stars were formed where they are currently observed, or they were formed away from the center but migrated to small radii, for example, due to gravitational scattering on other stars. Possibly, the young hot stars are responsible for the practically full ionization of hydrogen in the circumnuclear disk inside the inner 150~pc, while at larger distances (up to 1~kpc), molecular hydrogen,
whose density is almost two orders of magnitude higher than that of ionized hydrogen, accounts for only 4\% of the stellar mass \cite{96,97}. Some massive OB stars, apparently, were formed in collisions and coalescences of less massive stars. It is also interesting to note that in the distribution of stars around Sgr~A*, thin disks tilted by large angles to each other can be discerned, which could be due to a complex star
formation history of the central cluster. It is possible that the stellar density maximum does not coincide with Sgr~A* and is located $2''$ to the east \cite{109}. If this is indeed the case, the formation history of the central part of the Galaxy is more complicated.

The local relaxation time of the total orbital energy $E$ of a star due to binary collisions of stars in the central stellar cluster is \cite{111,112}
\begin{equation}
t_{\rm E}=\left(\frac{2}{3}\right)^{1/2}\frac{v^3}{3\pi G^2m^2n\Lambda},
 \label{tE}
\end{equation}
where $\Lambda=\ln(N/2)$ is the Coulomb logarithm, $N\sim10^7$ is the total number of stars in the system, $m$ is the mass of the star, and $n$ is the local number density of stars. Due to the high number density of stars, relaxation time (\ref{tE}) in the central cluster is significantly smaller than the galactic age,
$t_{E}\sim10^9$~years. This means that the dynamical evolution of the central stellar cluster caused by relaxation processes significantly affects its structure on a time scale exceeding the dynamical time $t_{\rm dyn}\sim10^3$~years. In self-gravitating stellar systems consisting of a large number of stars, the relaxation
time is significantly larger than the dynamical time: $t_{E}/t_{\rm dyn}\sim N/\ln N\gg1$ for $N\gg1$. On time scales much shorter than the relaxation time, i.\,e., for $t\ll t_{E}$, the motion of stars in such systems is collisionless, and the orbital energy of stars is conserved, $E=const$. Conversely, for $t\geq t_{E}$,
orbital energies of stars change in a diffuse way.

The dynamical evolution also depends on the gravitational interaction of stars with the central SMBH inside the
radius of its gravitational influence, $r_h=GM_h/v_{\rm vir}^2$. As a result, a stellar density peak can be formed near the BH due to the accumulation of stars in finite orbits (i.\,e., gravitationally connected with the BH) with energies $E<0$. The finite star reservoir is replenished due to infinite stars (those gravitationally disconnected from the BH) that transit to finite orbits due to gravitational interactions. The number density of infinite stars inside the BH influence radius $r_h$ scales as $n(r)\propto r^{-1/2}$ (see, e.\,g., \cite{113}).

A similar problem of the relaxation of a multi-charged ion in plasma with the identical interaction mechanism was first considered in 1964 by Gurevich \cite{114}, who calculated the distribution function of finite electrons around an ion as $f(E)\propto |E|^{1/4}$. This distribution function, which is a nonlinear isotropic solution of the stationary kinetic equation in the Fokker-Planck approximation, describes the diffusion of
electrons toward the multi-charged ion. This distribution function corresponds to the electron number density peak near the ion: $n(r)\propto r^{7/4}$. A similar problem for the number density of stars around an SMBH was considered by Bahcall and Wolf \cite{115}.

\begin{figure}
	\begin{center}
\includegraphics[angle=0,width=0.45\textwidth]{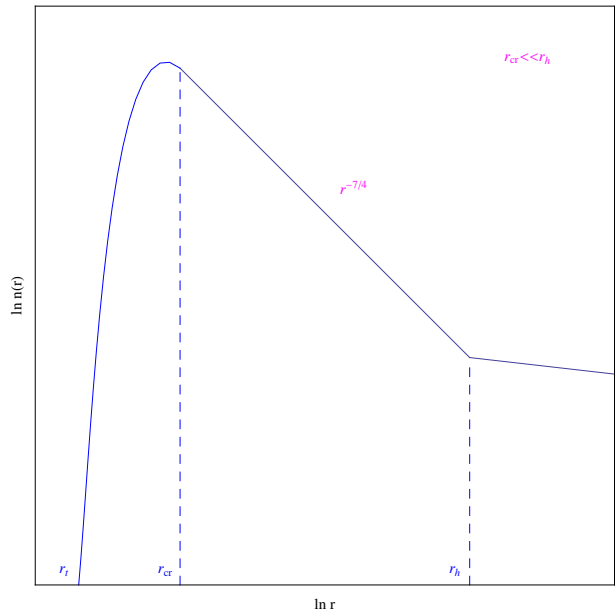}
	\end{center}
\caption{The peak in the finite-star number density $n(r)\propto r^{-7/4}$ near a
supermassive BH at $r_{\rm cr}\ll r_h$. The scale is arbitrary.}
	\label{fig1}
\end{figure}

\begin{figure}
	\begin{center}
\includegraphics[angle=0,width=0.45\textwidth]{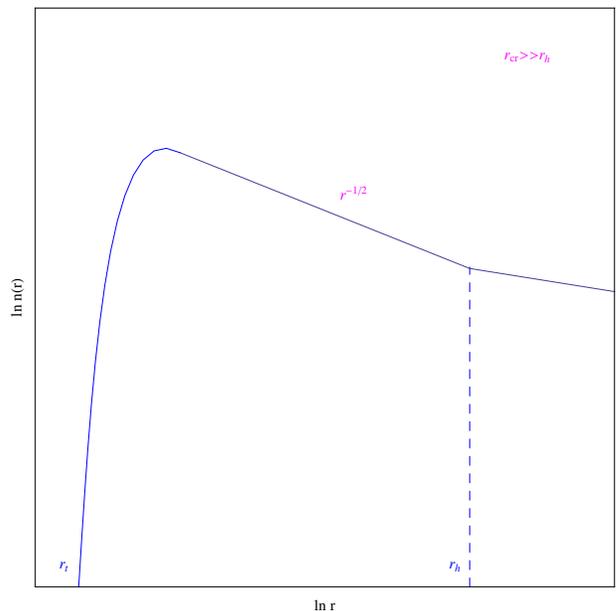}
	\end{center}
\caption{Number density distribution of infinite stars $n(r)\propto r^{-1/2}$ near an SMBH at $r_{\rm cr}\gg r_h$ in the absence of finite stars. The scale is arbitrary.}
	\label{fig2}
\end{figure}

At the Galactic center, the flux of solar-type stars onto a BH is determined by their entering the loss cone \cite{116,117}, in which a star with a mass $m_*$ and a radius $r_*$ is tidally disrupted by the BH \cite{118,119,120,121,122} by flying near it at a distance closer than the tidal radius $r_t\simeq2r_*(M_h/m_*)^{1/3}$. For solar-type stars, this radius exceeds the gravitational radius of Sgr~A*. Only white dwarfs and neutron stars are ``swallowed up'' by this BH as a whole without being tidally destroyed (see, e.\,g., \cite{6,123}). The tidal disruption of stars falling into the loss cone due to diffusion of their orbital angular momentum $J$ leads to an anisotropy of the distribution function at the orbital energies
of finite stars $E<E_{\rm cr}=-GM_hm_*/(2r_{\rm cr})$, where the critical radius $r_{\rm cr}$ under the condition $r_{\rm cr}<r_h$ is expressed as \cite{115}
\begin{equation}
r_{\rm cr} \simeq r_h\left(\frac{v_{\rm vir}r_tt_{\rm E}}{r_h^2}\right)^{4/9}\!\!.
 \label{cr}
\end{equation}
For $E<E_{\rm cr}$, the distribution function of finite stars in the gravitational field of a BH can be approximated as \cite{78,79,80,81,88}
\begin{equation}
f(E,J)\propto|E|^{1/4}\ln\frac{J}{J_{\rm min}},
 \label{fEJ}
\end{equation}
where $J_{\rm min}=(2GM_hr_t)^{1/2}m_*$ is the minimum angular momentum of stars at which they reach the tidal disruption radius $r_t$. The number density of stars attains a maximum at $r<r_{\rm cr}$ and tends to zero when approaching the tidal disruption radius $r_t$. At $r_{\rm cr}>r_h$, the density peak does not form because there is no accumulation of finite stars around the BH. Figures~\ref{fig1} and \ref{fig2} show the number density of stars around an SMBH at $r_{\rm cr}\gtrless r_h$. The above asymptotic formulas for stellar distribution around an SMBH are obtained with account for binary interactions from the corresponding solution of the Fokker-Planck equations. The binary interaction approximation remains partially valid for studies of dark matter particles inside the BH influence radius rh (see, e.\,g., \cite{82,84,85}). But determining the boundary conditions at the BH influence radius in order to normalize the dark matter particle density is complicated due to additional uncertainties related to the dark matter interaction with the Galactic
disk and the bulge \cite{83}. The dynamics and distribution of stars in the vicinity of an SMBH are also actively studied by numerical simulations, including Monte Carlo modelling (see, e.\,g., \cite{124,125,126,127,128,129}). Using numerical methods, it is possible in principle to study the
evolution of central galactic BHs on cosmological time scales.


\section{Formation and joint evolution of the Galaxy and its central black hole}

According to the hierarchical model of galaxy formation, which is currently commonly recognized, low-mass galaxies (protogalaxies) are first formed, which later merge to form large present-day galaxies. Because stellar disks are present in spiral galaxies, which were not destroyed in mergers, spiral galaxies must have been formed from single large density perturbations, and the mergers played a less significant role there than in ellipticals, at least at the late stages. Nevertheless, some protogalaxies could merge with our Galaxy. For example, the properties and composition of the bulge (central spherical component) of our and other spiral galaxies are very similar to those of small elliptical galaxies \cite{97}, and could be a remnant of one or several elliptical protogalaxies.

Observations in the 21~cm and 2.6~mm lines suggest the presence at the Galactic center of a gas disk 3~kpc in diameter and 200 pc thick, tilted to the Galactic plane by $22^\circ$ \cite{97}. This tilt could arise from mergers of stellar systems of protogalaxies with different orientations of their angular momentum. Many of the protogalaxies or satellite galaxies were tidally disrupted and mixed up with the Galactic matter. Several star
streams are known that most likely are remnants of satellite galaxies that merged recently with our Galaxy \cite{130}. It is also assumed that some of the Galactic globular clusters can be remnants of dwarf galaxies, whose outer parts were stripped by tidal gravitational forces \cite{131}.

To learn how the central SMBH was formed in the galaxy, it is necessary to understand at which stage the central BHs started forming in protogalaxies and how they subsequently merged. There are many reviews devoted to the formation of SMBHs in the galactic and pre-galactic epochs (see, e.\,g., \cite{132,133}), and we do not discuss these models here. We just mention several possible variants: gravitational collapse of supermassive stars and compact stellar clusters; multiple mergers of stellar-mass BHs (including those formed in explosions of the pre-galactic Population III stars); and the presence of primordial BHs with high masses that served as
seeds for galactic SMBHs \cite{134,135,136}.

After protogalaxies merge, central BHs can approach and coalesce. The coalescence mechanism of sufficiently massive BHs is generally well understood. It includes the dynamical friction that tends to move the remnant of a merging protogalaxy and its central BH toward the center of the new galaxy, and the orbital shrinking of a binary BH at late stages occurs due to gravitational wave emission. Low-mass BHs have no time to settle at the Galactic center over the galactic life time, and therefore they are likely to reside in the halo.
Because there is evidence that even small galaxies contain central BHs, the galactic SMBHs could form from the
coalescence of several less massive BHs. Possibly, these BHs originally resided in the protogalaxies that formed the galactic bulge.

The central dark matter density peak depends on how long ago the BH mergers stopped, because the density peaks
must be destroyed partially or completely in the mergers. However, it is currently impossible to restore the merger history of the central BH. Conversely, the possible annihilation signal allows density peak reconstruction and can be used to gain insight into the evolution history of the central galactic
supermassive BH.


\section{``Weighing'' of the invisible matter at the Galactic center}

Besides the visible stars, additional mass can exist inside the region between the BH event horizon and the influence radius in the form of both individual objects and diffuse matter. At first, this additional matter can include dim low-luminosity stars, which cannot be directly observed by telescopes. Because the stellar mass function increases toward low masses, these stars can significantly contribute to the total
density. At the Galactic center, there should be ``remnants'' of stellar evolution: stellar-mass BHs and neutron stars. Stellar-mass BHs, with relatively large masses, settle down closer to the Galactic center due to the mass segregation and form a compact central subsystem. Consequently, their contribution to the central density peak can be substantial \cite{137}. The contribution from neutron stars is unknown, but their
presence in the central density peak is very likely. An active pulsar was indeed discovered at a distance of 0.1~pc from the central BH \cite{138}; however, most of the neutron stars could be unavailable for observations. Constraints on the dark matter density at the Galactic center due to neutron star collapses into
BHs were discussed in \cite{139,140}.

Bounds on the number of stellar-mass BHs at the Galactic center can be derived from X-ray luminosity due to gas
accretion. The effect is strongest when a gas cloud falls onto the Galactic center. When this cloud flies across the central region, accretion onto usually ``quiet'' BHs and neutron stars can occur, and they can temporarily become bright in X-rays. This effect was discussed in \cite{141} with regard to the fall of the
G2 gas cloud onto the Galactic center. Since 2011, the G2 cloud has been observed approaching Sgr~A* with a velocity of several thousand km~s~${-1}$, stretched by tidal forces \cite{142}. However, no expected tidal disruption and subsequent accretion of gas happened at the G2 cloud periastron passage. Nevertheless, this can occur in coming years during the next approaches of G2 to Sgr~A*.

All invisible mass at the Galactic center must distort the total Newtonian potential $U_h=-G M_h/r$ of a point-like BH. As a result, orbits of the S0 stars are not closed and precess (see, e.\,g., \cite{143}). In one to two years, the degree of nonclosedness, i.\,e., the Newtonian orbital precession, will be
securely measured for the best studied S0-2 star. Thus, the total mass of dark matter within this stellar orbit with the characteristic size of 0.005~pc will be estimated. The non-relativistic precession of orbits of the high-velocity S0 stars can significantly exceed the corresponding relativistic precession (similar to the Mercury perihelion shift and the Lense-Thirring effect) even in the presence of a moderate amount of
dark matter in the Galactic center. 

The existence of high-velocity S0 stars offers a unique opportunity to reconstruct the gravitational potential and to measure the mass distribution in the Galactic center by measuring their orbits. For a power-law dark matter density
\begin{equation}
\rho(r)=\rho_h\left(\frac{r}{r_h}\right)^{-\beta},
\label{power}
\end{equation}
where $\rho_h$, $r_h$, and $\beta$ are constant parameters, the precession angle df of an elliptical orbit of an S0 star in one revolution around the central BH is expressed as \cite{144}
\begin{equation}
\delta\phi=-\frac{4\pi^2\rho_hr_h^\beta p^{3-\beta}}{(1-e)^{4-\beta}M_{\rm BH}}\,{_2F_1}\!\left(\!4-\beta,\frac{3}{2};3;-\frac{2e}{1-e}\right)\!,
\label{itog}
\end{equation}
where $_2F_1(a,b;c;z)$ is the hypergeometric function, $e$ is the orbital eccentricity, $p=a(1-e^2)$ is the orbital parameter, $a$ is the major semi-axis, and MBH is the BH mass. Nonrelativistic precession angle (\ref{itog}) is qualitatively consistent with numerical calculations of the precession  \cite{56,68,145,146,147,148}. The precession angle (\ref{itog}) for $e\ll1$ coincides with the corresponding expression obtained analytically by another method in \cite{149}. We note, however, that for large eccentricity
$e\simeq1$, the precession direction calculated in \cite{149} changes sign (becomes positive) and diverges in the limit as $e\to1$. Apparently, the method of calculations used in \cite{149} is applicable only for $e\ll1$, because the Newtonian precession angle $\delta\phi$ is always negative for other eccentricities and $\delta\phi=0$ for $e=1$. A multi-parameter fitting of orbits of S0 stars, as well as the estimate of possible additional diffuse mass \cite{23,24,25}, showed that the diffuse mass inside the S0-2 star orbit does
not exceed 3-4\% of the BH mass. The expected measurement of the nonrelativistic orbital precession of the S0-2 star will improve this limit by two to three orders of magnitude or will allow determining the dark matter mass inside this orbit.

The corresponding total mass of dark matter MDM within the sphere of a radius $r$ for power-law density profile (\ref{power}) has the form
\begin{equation}
M_{\rm DM}(r)=\frac{4\pi\rho_h r_h^\beta}{3-\beta}\left[r^{3-\beta}-R_{\rm min}^{3-\beta}\right],
\label{mrint}
\end{equation}
where $R_{\rm min}$ is the minimal radius up to which density profile (\ref{power}) can be extended. To calculate the orbital precession angle, we assume that $\beta<3$ and $R_{\rm min}$ is much smaller than
the orbital pericenter radius of the S0-2 star, $r_p=a(1-e)=0.585$~mpc. In this case, most of the dark matter
mass is confined near the apocenter, $r_a=a(1+e)=9.42$~mpc. For our analysis in what follows, it is convenient
to introduce the dark matter mass fraction inside the S0-2 star orbit: $\xi=[M_{\rm DM}(r_a)-M_{\rm DM}(r_p)]/M_{\rm BH}$.


\section{Dark matter annihilation at the Galactic center}

The physical nature and the composition of enigmatic dark matter are currently unknown. Most frequently, dark matter is assumed to be composed of still undetected elementary particles \cite{150,151,152}. Nor can it be ruled out that dark matter consists of primordial BHs \cite{153,154,155,156,157,158,159,160}, closed chiral cosmic strings (vortons) \cite{161,162,163,164,165,166,167}, nontopological solitons in the form of bosonic or fermionic Q-balls \cite{168,169} and Q-stars \cite{170}, massive gravitons \cite{171}, ultralight scalar fields \cite{172,173,174,175}, scalar stars made of dark energy in the form of complex scalar fields \cite{176,177}, multi-dimensional Kaluza-Klein particles (D-matter) and point-like defects \cite{178,179,180,181,182}, or is multi-component \cite{183,184,185,186,187}. Detailed dark matter models
include axions \cite{188,189,190,191,192,193,194,195,196}, massive sterile neutrinos \cite{197,198,199,200,201,202,203}, mirror particles \cite{204,205,206,207,208,209,210,211,212}, and the lightest stable supersymmetric particles, neutralinos \cite{34,213,214,215,216,217}. (Possible scenarios of neutralino annihilation at the Galactic center is discussed below.) As an alternative to the dark matter
phenomenon, different modified gravity models are also discussed (see, e.\,g.,  \cite{218,219,220,221,222,223,224,225} and the references therein).

In some low-background laboratories, experiments are underway that use nuclear recoil in an attempt to directly
detect dark matter particles crossing Earth. Positive detection has even been claimed (for example, in DAMA (Dark Matter experiment) at the Gran Sasso National Laboratory (Italy) \cite{226}), but no independent confirmation of this result has been obtained. The experimental search for signals from dark matter
particle annihilations (or decays) has been actively carried out, first of all, in gamma rays. The corresponding searches for the annihilation products are called indirect dark matter detection experiments.

The Galactic disk and the entire central part of the Galaxy contain many point-like and diffuse gamma-ray sources \cite{227}, which form a significant background for a possible dark matter annihilation signal. The Galaxy is filled with cosmic rays (energetic charged particle flows) accelerated in supernova remnants or other sources. Cosmic rays generate gamma-ray photons in interactions with the interstellar gas
(the Ginzburg-Syrovatskii secondary generation model \cite{228}). Clearly, background emission is mostly produced in the Galactic disk or at the Galactic center, which contain most of the gas. Due to the presence of numerous objects and the complex distribution of gas clouds, it is difficult to theoretically predict this gamma-ray background and to take it into account in data analysis.

By assuming the entire observed gamma-ray emission to be due to dark matter annihilation and the thermal mechanism of dark matter particle creation in the early Universe, it is possible to conclude that the upper bound of gamma-ray emission from dwarf spheroidal galaxies already excludes dark matter particles with masses $m<30$~GeV \cite{229}. However, attempts are being made to improve the background subtraction to find possible gamma-ray excess. For this, complex numerical models of the secondary generation of gamma-ray emission by cosmic rays are being constructed and careful data analyses of gamma-ray observations are being made. It is claimed in \cite{229} that dark matter particles masses $m < 100$~GeV are already excluded at the 68\% c.l. This result, of course, needs independent confirmation.

The analysis of data obtained by the first-generation gamma-ray telescope Compton-EGRET (Compton Gamma-Ray Observatory-Energetic Gamma Ray Experiment Telescope) revealed the presence of a gamma-ray excess at energies $\sim50-100$~GeV \cite{230}. This excess was thought to be well described by neutralino annihilation with masses $\sim100$~GeV. Later, more reliable Fermi-LAT observations did not confirm the excess at these energies. Nevertheless, an excess at lower energies of $1-3$~GeV was revealed in the Fermi-LAT data \cite{35,36,37,38,39,40,41,42,43,44}. This excess is observed from a spherical region a few degrees in radius around the Galactic center, and, importantly, the energy spectrum of the excess does not depend on the spatial position. The excess is well described by neutralino annihilation with the mass $m=35$~GeV and the thermal cross section $\langle\sigma v\rangle=1.7\times10^{-26}$~cm$^3$~s$^{-1}$ under the assumption that the dark matter density profile is described by the modified Navarro-Frenk-White law \begin{equation}
 \rho_{\rm H}(r)=
 \frac{\rho_{0}}{\left(r/d\right)^{\gamma}\left(1+r/d\right)^{3-\gamma}},
 \label{ghalonfw}
\end{equation}
where $\rho_{\rm H}(8.5\mbox{~kpc})=0.3$~GeV~cm$^{-3}$ and $d=20$~kpc. The observed gamma-ray excess is best fit by $\gamma\approx1.26$, while $\gamma=1$ in the standard profile.

The region$r\ll d$ is called the cusp \cite{231,232,233}. If additional density growth exists in the cusp region around the central BH inside the BH influence radius, it is called the spike. The BH gravitational field can redistribute the dark matter density profile to produce a spike.

The central density spike is mostly formed due to the central BH. The presence of a BH at the center of a dark
matter halo increases the central dark matter density inside the BH influence radius $r_h\sim GM_h/v_0\sim1.7$~pc, where $v_0\sim100$~km~s$^{-1}$ is the velocity dispersion inside the inner cusp. The density spike formed evolutionarily together with the BH and its properties are fundamentally dependent on the
BH formation history. If the BH was merging with other BHs, the dark matter density spike could be formed and destroyed several times. These perturbations decrease its density. Conversely, if the BH evolution was relatively quiet, i.\,e., if the BH was formed early and gradually increased due to slow accretion, then the dark matter density spike around the BH would also gradually increase to a high value \cite{234,235}. Such
a gradually formed density peak is called the ``adiabatic peak''. The results of calculations in \cite{234} suggest that the density distribution in the adiabatic peak is approximately power-law with the exponent $\beta=(9-2\gamma)/(4-\gamma)$, where $\gamma$ is the power-law exponent in (\ref{ghalonfw}). For example, for $\gamma=1.26$, we find $\beta=2.36$.

The maximum density $\rho_{\rm max}\sim m/(\langle\sigma v\rangle t_g)$ of dark matter in the spike is limited by the annihilation effect \cite{34,236}. The density increase stops at the inner radius where dark matter
has time to annihilate during the characteristic BH lifetime $t_g\sim10^{10}$~yrs. If annihilation is insignificant, the density growth continues to a few gravitational radii of the BH, where the capture of particles by the BH becomes important.

Scattering of dark matter particles by stars \cite{235} can also affect the dark matter density distribution in the spike to produce a universal dark matter density profile $\rho\propto r^{-3/2}$. Such a scattering also helps to fill the loss cone and increases the flux of particles onto the BH, leading to a gradual evolutionary increase in the BH mass \cite{83,85}. 

The angular size of the density peak is smaller than the angular resolution of the Fermi-LAT telescope. Therefore, the spike in gamma rays should be seen as a point-like source. Calculations of the annihilation signal from an adiabatic density spike \cite{237} show that in the case $\langle\sigma v\rangle=const$, even a
fairly small fraction of dark matter $\xi\sim10^{-5}$ within the S0-2 star orbit would produce the annihilation gamma-ray signal that could already be detected by Fermi-LAT. The absence of an annihilation signal (more precisely, the low upper bound on the possible signal) suggests a very small amount of
neutralino dark matter, with $\langle\sigma v\rangle=const$ at the Galactic center. But then this dark matter could not affect the orbits of stars at the level available for current observations. According to \cite{144}, the adiabatic spike could exist for particle with masses $m\sim10-100$~GeV only for a nonstandard annihilation cross section $\langle\sigma v\rangle\propto v^{-\eta}$, where $\eta\geq3$. Due to a high velocity dispersion near the BH, the annihilation cross section must decrease in this case. Then, in spite of the presence of a significant amount of dark matter in the density spike region, the annihilation signal would be weak due to the
small annihilation cross section. In this case, however, the large mass of dark matter can already provide detectable Newtonian precession of the S0-star orbits.

Thus, the search for a possible annihilation signal from the Galactic center is an independent additional method to determine the dark matter distribution. The gamma-ray excess from the Galactic center at $\sim1$~1 TeV observed by the HESS (High Energy Stereoscopic System) telescope could be explained by dark matter particle annihilation with the constraints from the stellar dynamics for a power-law dark matter density profile with a free power-law exponent \cite{238}. The possibility of imposing constraints on dark matter
annihilation from the stellar dynamics or precession of their orbits was also noted in \cite{147}. Figures~4 and 5 show the dark matter mass that is necessary to explain the possible gamma-ray excess from the
Galactic center according to the Fermi-LAT observations \cite{42,237}, assuming that dark matter mostly contributes to $\xi$ and consists of the supersymmetric neutralinos \cite{144}.

\begin{figure}
	\begin{center}
\includegraphics[angle=0,width=0.45\textwidth]{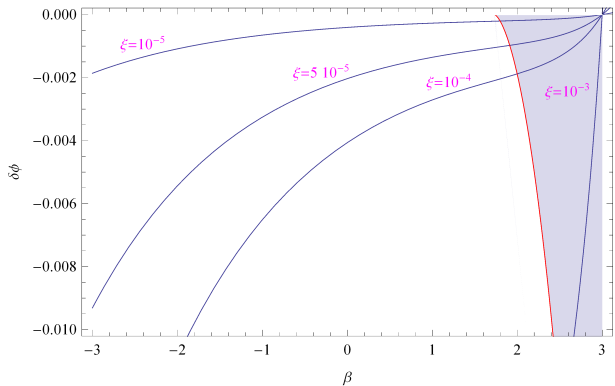}
	\end{center}
	\caption{Orbital precession angle $\delta\phi$, Eqn.~(\ref{itog}), as a function of the dark
matter radial density power-law exponent $\beta$ in (\ref{power}) for plausible dark matter mass fractions $\xi$ inside the S0-2 star orbit. The shaded area is excluded by dark matter (neutralino) annihilation constraints, if dark matter mostly contributes to $\xi$.}
	\label{figphi}
\end{figure}
\begin{figure}
	\begin{center}
\includegraphics[angle=0,width=0.45\textwidth]{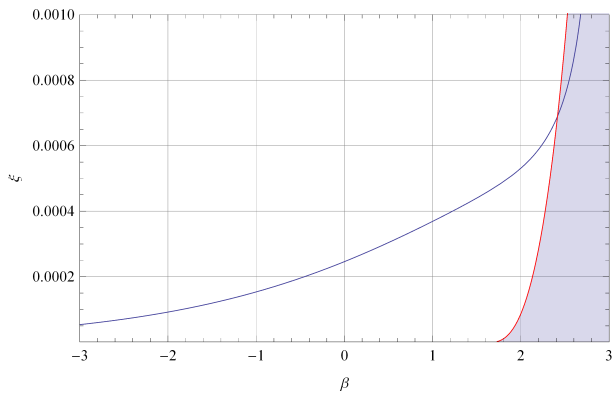}
	\end{center}
	\caption{Dark matter mass fraction $\xi$ as a function of the power-law exponent $\beta$ in the density profile (\ref{power}) for the precession angle $\delta\phi=0.01$. The shaded area is excluded by dark matter particle annihilation
constraints.}
	\label{figxi}
\end{figure}

If the annihilation cross section does have the standard thermal value, $\langle\sigma v\rangle\sim2\times10^{-26}$~cm$^3$~s$^{-1}$ (i.\,e. $\eta=0$), the observational gamma-ray constraints imply the absence of an adiabatic or sufficiently dense spike in the dark matter radial distribution. The absence of the adiabatic spike could be due to previous mergers of the central BH \cite{239} or tidal effects from the surrounding stars. Namely, stellar fly-by perturbs dark matter in the spike, such that the spike ``heats up''
and becomes less dense \cite{235}. Binary scatterings (due to self-interaction) of dark matter particles \cite{237} or a mismatch of the dark matter profile with the BH \cite{240} could be other reasons. The kinetic mixing during BH mergers \cite{239} and an off-center BH location \cite{240} yield a dark matter spike density
with $\beta=1/2$.

Searching for annihilation lines is one of the methods of signal detection above the noise level. Unfortunately, these lines are strongly suppressed in the most probable and simple dark matter models. Nevertheless, the detection of a possible line at $130$~keV from the Galactic center and some Galactic
clusters and unidentified Fermi-LAT sources was reported, which is intriguing but requires further verification \cite{241}.

We note that additional boosting of the annihilation signal is possible due to the dark matter clustering into low-scale self-gravitating clumps \cite{242,243,244,245,246,247,248,249,250,251}.

Apart from detection of gamma-ray emission, searching with giant neutrino telescopes (IceCube, ANTARES
(Astronomy with a Neutrino Telescope and Abyss environmental RESearch), Baikal-GCD (Baikal Gigaton Volume
Detector), etc.) for neutrino signals from the Galactic center, which can be generated in dark matter particle annihilation, is important \cite{234,252,253}.

The INTEGRAL (INTErnational Gamma-Ray Astrophysical laboratory) telescope detected an electron-positron
511~keV annihilation line signal from the $8^\circ$ region around the Galactic center \cite{254,255}, whose origin remains enigmatic. At the dawn of gamma-ray astronomy, I.S.~Shklovsky and L.M.~Ozernoi assumed that supernova explosions and young crab-like pulsars could be the source of positrons in the Galactic center (the 511~keV annihilation line is also observed in the Crab Nebula, which could simultaneously produce the
observed helium excess \cite{97}). However, it cannot be ruled out that positrons are produced by annihilations (or decays) of heavier dark matter particles.

Essentially, the search for signals from dark matter particle annihilation is a promising tool to probe the dark matter distribution at the Galactic center, if dark matter can annihilate and the annihilation products can be detected. The preliminary intriguing claims of the detection of a possible annihilation signal must be carefully checked. In particular, more detailed calculations of the observed astrophysical gamma-ray background are required.


\section{Spin measurement of the central Galactic black hole}
\label{QPO}

The axially symmetric stationary metric of a rotating Kerr black hole is determined by its mass $M_h$ and specific angular momentum$J/M_h=(GM_h/c)a$, where $a$ is the dimensionless spin parameter (or simply the spin) of the BH, ranging in the interval $0\leq a\leq1$. The gravitational field of such a BH in coordinates $(t,r,\theta,\phi)$ depends on the radius $r$ and the latitude angle $\theta$. The angular momentum (spin) of the BH is a very important parameter determining observational appearances of the accretion flow, generation of relativistic flows (jets), and particle acceleration near the BH. Future observations of the shadow around Sgr~A* with the Einstein Horizon Telescope will allow independent measurements of the mass and spin of the BH at the Galactic center. The BH spin can also be estimated from specific features of the emission and polarization from the innermost stable orbits of the accretion disk \cite{256,257,258}, from the analysis of continuum emission \cite{259,260,261,262}, by the specific form of emission lines \cite{263,264,265,266}, or by using the correlation between the power of jets and the BH spin \cite{267}.

The interpretation of interferometric observations at $1.3$~mm using the model of relativistic accretion onto a rotating BH \cite{259,260,261} leads (albeit with large uncertainties) to the most probable value of the Sgr~A* spin: $a=0+0.2+0.4$, where errors are shown in parentheses \cite{268}.

Quasiperiodic oscillations observed during rare outbursts from Sgr~A* with a period of 19~min in the IR band \cite{76} and about 11.5~min and 19~min in the X-ray band \cite{77}, mentioned in Section~\ref{observations}, contain important information on the parameters (mass and spin) of the BH and properties of the surrounding accretion flow. Quasi-periodic oscillations are usually related to resonances in accretion disks \cite{20,76,77,269,270,271}. The uncertain model-dependent structure of accretion disks is a weak point of this interpretation. Apparently, the resonance models of quasiperiodic oscillations can be applied to BHs in active galactic nuclei and binary stellar systems with high accretion rates. In \cite{271}, the detection of
quasiperiodic oscillations in radio emission from Sgr~A* is reported, with mean periods 16.8, 22.2, 31.4, and 56.4~min, which in the resonance oscillation models yield the Sgr~A* spin $a= 0.44\pm0.08$.

\begin{figure}
	\begin{center}
\includegraphics[angle=0,width=0.45\textwidth]{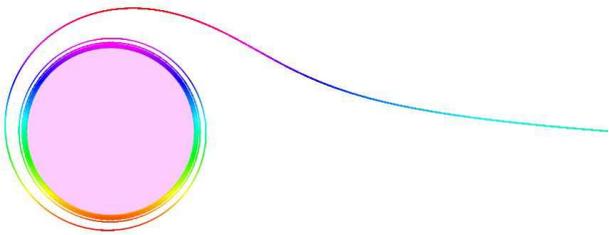}
	\end{center}
	\caption{Example of a numerical solution of the equation of motion for a photon falling with a zero impact parameter onto an extremal BH ($a=1$) in the equatorial plane. The photon approaching the BH is dragged into rotation and winds up many times on the event horizon in the BH spin direction with the azimuthal frequency $\Omega_{\rm h}$, Eqn~(\ref{OmegaH}). The inner region of the BH is shown in gray.}
	\label{photon}
\end{figure}

At low accretion rates typical for the modern state of the Sgr~A* SMBH, the accretion flow can be transparent down to the BH event horizon. Under such conditions, quasiperiodic oscillations could be produced by bright spots in the accretion flow, with frequencies being independent of the accretion model and fully determined by the BH gravitational field. In accretion disks, there are three characteristic frequencies: the frequency of rotation of the BH horizon 
\begin{equation}
 \Omega_{\rm h}=\frac{2\pi}{T_{\rm h}}=\frac{a}{2(1+\sqrt{1-a^2})}\frac{c^3}{GM_h},
 \label{OmegaH}
 \end{equation}
the frequency of azimuthal circular oscillations in the
equatorial plane
\begin{equation}
 \Omega_{\varphi}=\frac{2\pi}{T_{\varphi}}=\frac{1}{a+x^{3/2}}\frac{c^3}{GM_h}.
 \label{Omegaphicirc}
 \end{equation}
and the frequency of the latitudinal precession in a thin
accretion disk
\begin{equation}
\Omega_{\theta}=\frac{2\pi}{T_{\theta}}
=\frac{\sqrt{x^2-4ax^{1/2}+3a^2}}{x(a+x^{3/2})}\frac{c^3}{GM_h},
 \label{OmegathetaQ0}
\end{equation}
where $x=r/r_g$ is the dimensionless radial coordinate, $r_g=GM_h/c^2$, and $T_{\rm h}$, $T_{\varphi}$, and $T_{\theta}$ are the corresponding oscillation periods detected by a remote observer. The event horizon frequency $\Omega_{\rm h}$ does not depend on the latitude angle $\theta$; hence, the horizon rotates as a solid body \cite{272}. All particles approaching the horizon, including photons, are unavoidably dragged into rotation with the event horizon frequency $\Omega_{\rm h}$. The frequencies $\Omega_{\varphi}$ and $\Omega_{\theta}$ depend only on the radial coordinate, but the related oscillation signal is observed with
maximum intensity from the region of highest accretion energy release. This region is located near the last marginally stable circular orbits $x=x_{\rm ms}$ (see the definition, e.\,g., in \cite{257,273}).

A rotating BH drags all approaching particles into rotation. This leads to the appearance of the ergosphere,
inside which the azimuthal rotation opposite to that of the BH is impossible. The outer boundary of the ergosphere in the Boyer-Lindquist coordinates \cite{274} is determined from the equation $g_{00}=0$, which has the solution $x=x_{\rm ES}(\theta)=1+\sqrt{1+a^2\cos^2\theta}$. The inner boundary of the ergosphere
coincides with the BH horizon, $x_+=1+\sqrt{1-a^2}$.

Figure~\ref{photon} shows the trajectory of a photon falling with a zero impact parameter onto a rotating BH. The photon approaching the BH is dragged into the BH rotation and circles around the BH horizon many times in the BH rotation direction with the horizon rotation frequency Oh from (\ref{OmegaH}).

\begin{figure}
	\begin{center}
\includegraphics[angle=0,width=0.45\textwidth]{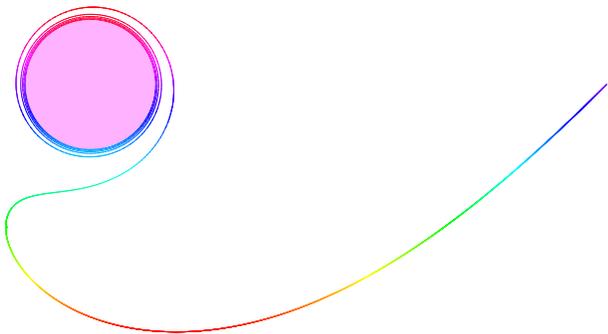}
	\end{center}
	\caption{Example of a numerical solution of the equation of motion for a photon falling with the negative impact parameter $b=-6.5r_g$ in the equatorial plane onto an extremal BH ($a=1$), which inverses its azimuthal rotation by approaching the BH. The photon winds up many times on the event horizon with the azimuthal frequency $\Omega_{\rm h}$, Eqn~(\ref{OmegaH}). The inner region of the BH is shown in gray.}
	\label{photon2}
\end{figure}

The inevitability of dragging the infalling particles into BH rotation is most prominent for particles with a negative azimuthal angular momentum relative to the BH spin. Figure~\ref{photon2} shows the trajectory of a photon falling with a negative impact parameter onto a rotating BH.

The logarithmic divergence of the time of approaching the BH horizon for the remote observer is a well-known feature of the fall onto a nonrotating BH. This time turns out to be finite with account for the inevitable back reaction of the falling particle on the BH, which leads to the BH mass increase. Qualitatively taking this effect into account, by analogy with white hole evolution \cite{275}, leads not to divergent but to a
finite logarithmic term for the corresponding fall time onto a BH:
\begin{equation}
t_{\rm obs}\sim\frac{r_g}{c}\ln\frac{M_h}{E},
 \label{finite}
\end{equation}
where $E$ is the total energy of the infalling particle.

An amazing additional feature of trajectories of test particles (including photons) falling onto a BH is the infinite number of rotations they make near the horizon from the remote observer's standpoint \cite{50}. With account for the back reaction of the infalling particle, in full analogy with the case of a nonrotating BH in (\ref{finite}), the number of rotations of a particle falling onto a rotating BH is also finite, $\sim\ln(M_h/E)$.

\begin{figure}
	\begin{center}
\includegraphics[angle=0,width=0.45\textwidth]{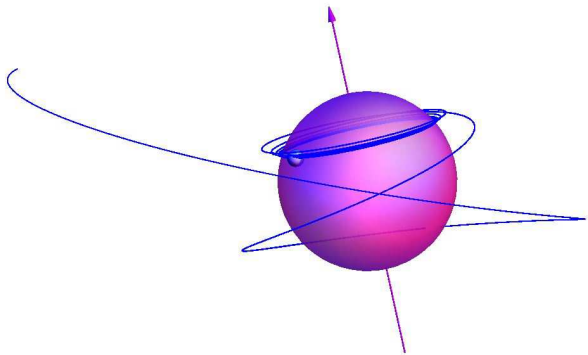}
	\end{center}
	\caption{Example of a numerical solution of the equation of motion (geodesic equation) for the $3D$ trajectory of a test particle (small ball) falling onto a rotating BH with the spin $a=0.998$ in the Boyer-Lindquist coordinates \cite{274}. In approaching the BH, the particle winds up many times at a fixed latitude on the BH event horizon with the angular frequency $\Omega_{\rm h}$, Eqn~(\ref{OmegaH}).}
	\label{infall}
\end{figure}

Figure~\ref{infall} presents the 3D trajectory of a particle falling into a rotating BH that winds up around the BH event horizon with the azimuthal frequency $\Omega_{\rm h}$, Eqn~(\ref{OmegaH}).

Any source of emission, for example, a hot plasma blob or a bright spot in the accretion disk, is registered by the remote observer in the relativistic ``synchrotron mode'' as short bursts collimated within a narrow solid angle \cite{273,276,277,278}, which repeat quasiperiodically with a frequency close to that of the BH horizon rotation $\Omega_{\rm h}$. It is this modulation of the signal from many clumps or bright spots in accretion disks that can be detected from the BH when the surrounding plasma is transparent to emission down to the BH event horizon.

\begin{figure}
	\begin{center}
\includegraphics[angle=0,width=0.45\textwidth]{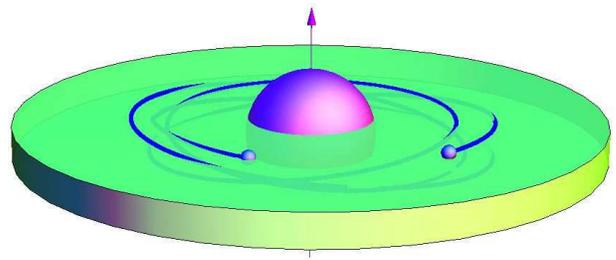}
	\end{center}
	\caption{Result of a numerical solution for the latitudinal precession of a bright spot or plasma blob (small ball) in a semi-transparent accretion disk around a BH with the spin $a=0.65$ \cite{33} leading to observed oscillations with the frequency $\Omega_{\theta}$, Eqn~(\ref{OmegathetaQ0}).}
	\label{fig6}
\end{figure}

The latitudinal precession frequency $\Omega_{\theta}$ in Eqn~(\ref{OmegathetaQ0}) can also explain quasiperiodic oscillations. Hot plasma blobs precess along the latitude angle in the accretion disk and, if the disk is opaque or semi-transparent, these blobs can be observable only during their ``rising'' to the disk surface turned to the observer. These events can be observed as quasiperiodic oscillations with the frequency $\Omega_{\theta}$. Figure~\ref{fig6} shows the latitudinal precession of a bright spot or plasma blob in a semi-transparent accretion disk that leads to the observed oscillations with the frequency $\Omega_{\theta}$ in (\ref{OmegathetaQ0}).

We note that unlike the sharp modulation of emission from an accretion disk with frequencies $\Omega_{\rm h}$ and $\Omega_{\theta}$, the modulation with the frequency $\Omega_{\varphi}$ is small. Indeed, the rotation of hot blobs in the accretion disk should not strongly modulate its emission with the frequency Oj due to the small ratio of the accretion disk size to the distance to the BH. Modulation with the frequency $\Omega_{\varphi}$ can be noticeable only in a strongly nonstationary accretion flow. These arguments justify the method of two characteristic frequencies to interpret quasiperiodic oscillations from the supermassive BH at the Galactic center observed in the IR and X-ray bands \cite{76,77}.

For the observed frequencies, the joint solution of Eqns~(\ref{OmegaH}) and (\ref{OmegathetaQ0}) at $x=x_{\rm ms}$ for the Sgr~A* mass $M_h$ and its spin $a$ uniquely determines their most precise present values
$M_h= (4.2\pm0.2)\, 10^6M_\odot$ and $a=0.65\pm0.05$ \cite{33}. Here, the period of 11.5~min of the observed quasiperiodic oscillations is identified with the rotation of the BH horizon, and the 19-min oscillations are identified with latitudinal oscillations of hot spots in the accretion disk. These oscillations provide important evidence of the presence of a BH at the Galactic center. The validity of this interpretation is
justified by the independent mass measurement of Sgr~A*, which is exactly the same as derived from observations of S0 stars \cite{33}. We note that different methods \cite{33,271} yield different spins a of the Sgr~A* BH. A new independent spin estimation of Sgr~A* is needed, e.\,g., from the Einstein Horizon Telescope observations of the BH event horizon.


\section{Conclusion}

The natural laboratory at the Galactic center offers great opportunities to study many physical processes near the supermassive black hole Sgr~A*. The shadow of this black hole is expected to be measured in the nearest future, which will test GR in the strong-field regime. This experiment also opens the avenue to check various modifications and generalizations of GR.

In the astronomically small volume around the Galactic center, processes from almost all fields of modern physics occur. For example, mechanics and Einstein's GR are presented by the supermassive BH Sgr~A*, as are the orbital motion of stars around the BH and gas accretion. Thermodynamics, hydrodynamics, and electrodynamics are used to describe plasma in stars and hot gas. Atomic and molecular physics, together with quantum mechanics, are needed to understand line formation in atomic and molecular transitions, and nuclear physics is required to explain processes in stars and to construct dynamical and evolutionary models of
the Galactic center. The physics of elementary particles (high-energy physics) is used to explain the generation of gamma-ray emission, for example, by dark matter particle annihilation around Sgr~A*. Clearly, modern physics is largely involved in any astrophysical object. In this review, we attempted to illustrate the diversity of physical phenomena in a very intriguing region of the observable Universe, the center of the Milky
Way galaxy.


\section*{Acknowledgments}

The authors are grateful to V. A.~Berezin, V. S.~Beskin, E. A.~Vasiliev, M. I.~Zelnikov, K. P.~Zybin, Ya. N.~Istomin, and S. G.~Rubin for the useful discussions and critical notes. The work was supported by the Program of the Division of Physical Sciences of RAS OFN-17 and the RFBR grants 13-02-00257 and NSh~3110.2014.2.


\end{document}